\documentclass[twoside]{article}
\usepackage{fleqn,espcrc2}

\usepackage{graphicx}
\usepackage[figuresright]{rotating}

\newcommand{\AmS}{{\protect\the\textfont2
  A\kern-.1667em\lower.5ex\hbox{M}\kern-.125emS}}

\title{Stripe ordering and two-gap model for underdoped cuprates}

\author{
C. Castellani, C. Di Castro, M. Grilli, A. Perali\address{INFM 
and Dipartimento di Fisica,
Universit\`a di Roma ``La Sapienza'' \\ 
Piazzale Aldo Moro, 2 - 00185 - Rome, Italy}
}
       
\begin{document}

\begin{abstract}
\vspace{1pc}
The evidence of edge-gaps around the M-points 
in the metallic state of underdoped cuprates
has triggered a very active debate on their origin. We first
consider the possibility that this spectroscopic feature results
from a quasi-static charge ordering taking place in the
underdoped regime. It comes out 
that to explain the coexistence of gaps
and arcs on the Fermi surface the charge modulation should be
in an eggbox form. In the lack of evidences for that, we then 
investigate the local pairing induced by charge-stripe 
fluctuations. A proper description of the strong anisotropy of both
the interactions and the Fermi velocities requires a two-gap model
for pairing. We find that a gap due to incoherent pairing forms
near the M-points, while coherence is established  by
the stiffness of the pairing near the nodal points.
The model allows for a 
continuos evolution from a pure BCS pairing 
(over- and optimally doped regime) to a mixed boson-fermion
model (heavily underdoped regime).  
\end{abstract}

\maketitle

\section{INTRODUCTION}

The opening of a pseudogap below a strong doping $(\delta)$ dependent 
crossover temperature $T^*(\delta)$, already above the
superconducting critical temperature $T_{c}(\delta)$, 
is an intriguing feature of the underdoped cuprates  \cite{timusk}.
One possibility is that the pseudogap arises from strong
scattering processes in the particle-hole channel. In this
framework, spin fluctuations have been considered 
\cite{chubukov}. A similar outcome 
would arise from scattering by charge fluctuations
near a charge instability for stripe
formation \cite{prl95,prb96,jpcs98,noiqui}.
It was indeed suggested that the tendency to spatial charge order
(which evolves into a mixed spin-charge stripe phase
by lowering the doping) gives rise to an instability
line ending in a Quantum Critical Point (QCP) at $T=0$ near optimal
doping. By approaching the instability line,
the quasi-critical stripe fluctuations
affect the states on the FS in a quite anisotropic way.
The fermionic states around the M points [i.e.
$(\pm \pi, 0),(0,\pm \pi)$] interact strongly
(these are the so-called ``hot spots''), while the states
around the $\Gamma$-$X$ or $\Gamma$-$Y$ diagonals are
weakly interacting (``cold spots'').
Well below the instability line, deep in the underdoped phase, 
a local (quasi-static) stripe order takes place, which can
strongly affect the spectral distribution of the quasiparticles
\cite{seibold}.
This mechanism for  pseudogap formation will be discussed in the 
next section. On the other hand a second possibility is that
the pseudogap arises from pairing in the
particle-particle channel with
$T^*$ being a mean-field-like temperature where 
electrons start to form local pairs without phase coherence.
Lowering the temperature, the phase coherence 
and hence superconductivity are established at the
critical temperature $T_{c}$. In this context it is still
debated whether the superconducting transition
is intermediate between a BCS transition and a Bose-Einstein
condensation of preformed pairs \cite{randeria}, 
or is due to a more intricate 
interplay between preformed bosonic pairs and fermions 
\cite{ranninger,GIL}.
Within the Stripe-Quantum-Critical-Point (Stripe-QCP) scenario
a two-gap model can be considered, where strongly paired fermionic states 
can coexist and interplay with weakly coupled pairs in different regions
of the Fermi surface (FS). This is a natural
description of the cuprates near the instability line, when quasiparticles 
have very different dispersions and effective interactions in different
regions of the Fermi surface. In this case the strong momentum dependent
singular scattering arising in the proximity of the stripe
instability acts in the particle-particle channel and 
leads to tightly bound (strongly phase fluctuating)
pairs near the hot spots (close to the M points) 
coexisting with weakly interacting
quasiparticles near the $\Gamma$-$X$ or $\Gamma$-$Y$ directions.
In this case the pseudogap will find a natural interpretation
in terms of the incoherent Cooper-pair formation around the
M-points. We will discuss this model in Section 3.

\section{CHARGE ORDERING IN THE UNDERDOPED CUPRATES}

Within the Stripe-QCP scenario the underdoped region
of the cuprates corresponds to the nearly ordered stripe
phase where a local (quasi-static) charge ordering
takes place. In this framework, a mean-field analysis 
will likely capture the main qualitative features of
the (locally) charge-ordered phase. In particular, in Ref.
\cite{seibold} the spectral properties of an incommensurate
charge-density-wave system were investigated within a standard
Hartree-Fock approach. Both a purely one-dimensional 
ordering with an order parameter 
$<\rho_{q}>^{1D}=
\sum_{n}<\rho_{q}>\delta_{q,nq_{c}^{x,y}}$
and a two-dimensional ``eggbox'' structure with
 $<\rho_{q}>^{egg}=\sum_{n}<\rho_{q}>\lbrack \delta_{q,nq_{c}^{x}}
 + \delta_{q,nq_{c}^{y}}\rbrack$
were considered. The effective density-density interaction and
 the critical wavevectors ${\bf q}_c$ were
 derived from microscopic calculations based on the frustrated
phase-separation mechanism for an Hubbard-Holstein model with
long-range Coulomb interaction.
A first generic outcome was that charge ordering tends
to substantially enhance the van Hove singularities near the
M points. Pseudogaps are also formed. However
the charge ordering modulated along one single direction,
$x$ or $y$, naturally produces pseudogaps along one direction only. 
The  superposition of two one-dimensional CDW modulated along
perpendicular directions simply fills up the pseudogap
in all four M-points. On the other hand, in the eggbox case 
a leading-edge gap arises near these points,
leaving finite arcs of the Fermi surface
gapless. This latter non-trivial feature might account
for recent ARPES results \cite{norman}, but does not seem to be robust
under the disordering of the eggbox modulation \cite{seibold}.
Therefore, although the particle-hole scattering definitely seems to 
affect the electronic spectra, particularly around the M points, additional 
mechanisms, likely related to the particle-particle pairing discussed
in the next section, seem to be in order to fully account for
the observed pseudogap. On the contrary, in the specific 
cases where commensuration effects couple the 
stripes to the underlying lattice structure,
charge ordering becomes particularly strong and it can alone 
open a full insulating gap. This is the case of
various hole-doped compounds at doping $\delta=1/8$,
where stripes in the (1,0) or (0,1) directions with half an hole in excess
per site (half-filled vertical stripes, HFVS) 
are observed \cite{tranquada}. 
In this regard it was recently established
\cite{seibold2} within a joynt (slave-boson)-(unrestricted Hartree-Fock) 
approach, that both a proper
treatment of the strong local hole-hole interaction
and the presence of a sizable long-range Coulomb 
repulsion are needed to obtain a ground state with HFVS.
The same conclusion was reached within a realistic three-band extended
Hubbard model \cite{sadori}, where also
the electron doped case was investigated. For electron doping
the most stable configuration was predicted to have
the stripes along the 
(1,1) or (1,-1) directions with one additional electron per site (filled
diagonal stripes).

\section{THE TWO-GAP MODEL}

As discussed above, the formation of 
a pseudogap in the metallic underdoped phase is likely
related to the strongly anisotropic attractive potential 
arising in the proximity of the instability.
In order to capture the relevant physical effects of the anisotropy
of both the pairing interaction
and the Fermi velocity, we introduce a simpliflied two-band model
for the cuprates \cite{twogap}

We describe the quasiparticle arcs of FS 
about the nodal points by a free electron
band  (labelled below by the index 1)
with a large Fermi velocity $v_{F1}=k_{F1}/m_1$ and the 
hot states about the M points with a second free electron band, 
displaced in momentum and by an energy $\varepsilon_0$
from the first, with a small
$v_{F2}=k_{F2}/m_2$. The energy $\varepsilon_0$ is
introduced to allow the chemical potential to cross
both bands: $E_{F1}=\varepsilon_0+E_{F2}$.
Moreover, since 
our main interest is the interplay between strongly and weakly coupled
pairs irrespectively of their symmetry, for simplicity we
assume a s-wave pairing interaction \cite{notaswave}.

The model Hamiltonian for pairing in the two-band system is
then

\begin{equation}
H_{pair}=
-\sum_{kk^{\prime}p\alpha\beta}g_{\alpha,\beta}
c^+_{k^{\prime}+p\uparrow \beta}c^+_{-k^{\prime}\downarrow \beta}
c_{-k\downarrow \alpha}c_{k+p\uparrow \alpha}
\label{hamiltonian}
\end{equation}
where $\alpha$ and $\beta$ run over the band 
indeces 1 and 2.
We also introduce a BCS-like energy cutoff $\omega_0$ to regularize 
the pairing interaction.
The $2\times 2$ scattering matrix $\hat g$ accounts for 
the strongly $q$-dependent effective interaction in the p-p
channel of the original
single-band system. Its elements $g_{ij}$ are
the different coupling constants which couple the electrons 
in the p-p channel within the
same band ($g_{11}$ and $g_{22}$) and between different bands 
($g_{12}=g_{21}$). 
The ladder equation for the superconducting fluctuation
propagator is given by
$\hat{L}=\hat{g}+\hat{g}\hat{\Pi}\hat{L}$, where
the  particle-particle bubble operator for the two-band spectrum
has a diagonal $2\times 2$ matrix form with elements
$\Pi_{11}({\bf q})$ and $\Pi_{22}({\bf q})$. The resulting 
fluctuation propagator is given by
\begin{equation}
\hat{L}({\bf q})=
\left(\begin{array}{cc}
\tilde{g}_{11}-\Pi_{11}({\bf q})&\tilde{g}_{12}\\
\tilde{g}_{12}&\tilde{g}_{22}-\Pi_{22}({\bf q})\\
\end{array}\right)^{-1}
\label{propmatrix}
\end{equation} 
where we have defined $\tilde{g}_{ij}\equiv (\hat{g}^{-1})_{ij}$.
It turns out useful to define 
the temperatures $T_{c1}^0$ and $T_{c2}^0$ as
$\tilde{g}_{11}-\Pi_{11}(0,T)\equiv \rho_1 \ln \frac{T}{T_{c1}^0}$,
$\tilde{g}_{22}-\Pi_{22}(0,T)\equiv \rho_2 \ln \frac{T}{T_{c2}^0}$,
where $\rho_i=m_i/(2\pi)(i=1,2)$ 
is the density of states of the $i$-th band.
To emulate the hot and cold points we assume 
$g_{22}>>g_{11}\simeq g_{12}$.
In this limit $T_{c1}^0$ and $T_{c2}^0$ (with $T_{c2}^0 \gg T_{c1}^0$)
give the two BCS critical temperatures for the two decoupled bands
(i.e. for $g_{12}=0$). For the coupled system ($g_{12}\ne 0$)
the mean-field BCS superconducting critical temperature 
$T_c^0$ is defined by
the equation $\mbox{det}\hat{L}^{-1}({\bf q}=0, T_c^{0})=0$.
We then obtain $T_c^{0}>T_{c2}^{0}$ given by 
\begin{equation}
T_c^{0}=\sqrt{T_{c1}^{0}T_{c2}^{0}}\exp\left[{\frac{1}{2}
\sqrt{\ln^2\left(\frac{T_{c2}^{0}}{T_{c1}^{0}}\right)+
\frac{4\tilde{g}_{12}^2}{\rho_1\rho_2}}}\right].
\label{tcmf}
\end{equation} 
The role of fluctuations can be investigated within a
standard Ginzburg-Landau (GL) scheme, 
when both $g_{22}< E_{F2}$ and $\omega_0< E_{F2}$.
We will assume that the chemical potential is not affected 
significantly by pairing and that fluctuations from the 
BCS result are not too strong.
The relevance of the space fluctuations of the order parameter
is assessed by  the gradient term coefficient
$\eta$, which provides the momentum dependence of the propagator.
This calculation requires the expansion of the fluctuation 
propagator in Eq.(\ref{propmatrix}) in terms of $q$. 
In particular the expansion of the particle-particle 
bubbles reads
$\Pi_{11}(q)\simeq\Pi_{11}(0)-\rho_1\eta_1 q^2$ and 
$\Pi_{22}(q)\simeq\Pi_{22}(0)-\rho_2\eta_2 q^2$. 
Here $\eta_i(i=1,2)$ is given by 
$\eta_i=(7\zeta(3)/32\pi^2)v_{Fi}^2/T^2$, with  $\eta_1\gg\eta_2$.
We obtain $\eta=\alpha_1\eta_1+\alpha_2\eta_2$
with
\begin{equation}
\frac{\alpha_1}{\alpha_2}=\frac{\tilde{g}^2_{12}}
{\rho_1\rho_2\ln^2(T_c^{0}/T_{c1}^{0})}, 
\label{etaeff}
\end{equation}
and $\alpha_1+\alpha_2=1$.
The presence of a fraction of electrons with a large 
$\eta_1$ increases the stiffness 
 of the whole electronic system $\eta$ with respect to $\eta_2$. 
However when the mean-field critical temperature $T_c^{0}$ is much 
larger than $T_{c1}^{0}$ the correction to $\eta_2$ due to the
interband coupling is small. At the same time the Ginzburg number 
is large implying a sizable
mass correction $\delta\epsilon (T)$ 
due to fluctuations to the ``mass'' 
$\epsilon (T)$ of the bare propagator $\hat{L}({\bf q})$. 
The renormalized critical temperature $T_c^{r}$, given by the equation
$\epsilon(T_c^r)+\delta\epsilon (T_c^r)=0$,
is lower than $T_c^{0}$ \cite{notaren}.
We find that the renormalized gradient term coefficient 
$\eta^r$, in the presence of the mass correction
is still given by Eq. (\ref{etaeff})
with $T_c^0$ replaced by $T_c^r$. Therefore, 
while mass renormalizations of the
fluctuation propagator tend to lower $T_c$, at the same time,
this increases the gradient term coefficient $\eta$ 
by increasing the coupling to $\eta_1$. As a consequence
the effective Ginzburg number is reduced and 
the system is stabilized with 
respect to fluctuations allowing for a coherent superconducting phase
even in the extreme limit $\eta_2=0$.
Within the GL approach we associate the temperature
 $T_{c}^{0}\sim T_{c2}^{0}$ to
the crossover temperature $T^*$ and $T_c^r$ to the superconducting 
critical temperature $T_{c}$ of the whole system. 

Within the Stripe-QCP scenario the coupling 
$g_{22}$ is related to the singular part 
of the effective interaction mediated by the stripe fluctuations.
$g_{22}$ is the most doping dependent coupling and attains 
its largest value in the underdoped regime.
$g_{11}$ and $g_{12}$ are instead less affected by doping.
In the region of validity of the GL approach,
the explicit calculations show that $r(\delta)\equiv \frac{T^*-T_c}{T^*}
\simeq\frac{T^0_c-T^r_c}{T^0_c}$ is increasing by increasing 
$g_{22}$, {\em i.e.}, by decreasing doping. For small values of $r$
we find that both $T^*$ and $T_c$ increases. 
This regime corresponds to the overdoped and optimally doped region.
For $r\sim 0.25\div 0.5$, $T_c$ is instead decreasing while 
$T^*$ is always increasing by decreasing doping. The large values
of $r$, which are attained in the underdoped region show that we
are reaching the limit of validity of our GL approach. 
We think however that the behavior of the
bifurcation between $T^*$ and $T_c$ represents correctly 
the physics of the pseudogap phase, at least qualitatively, while a
quantitative description would require a more sophisticated approach
like a RG analysis.

In the very low doping regime, where $T^*$ has increased strongly,
the value of $g_{22}$ can be so large 
to drive the system in a strong coupling regime for the
fermions in band 2 ($g_{22}> E_{F2}$).
In this case the chemical potential 
is pulled below the bottom of the band 2. 
The GL scheme must 
be abandoned and, in the limit of tightly bound 2-2 pairs,
the propagator $L_{22}({\bf q})$ assumes the form of a single pole
for a bosonic particle (similarly to the single-band strong-coupling
problem \cite{hausmann}).
The critical temperature of the system is 
again obtained by the vanishing 
of the inverse of $\det {\hat L}^{-1}$ at $q=0$
where, however, the chemical potential is now self-consistently evaluated
including the selfenergy corrections to the Green 
function in band 2 and the fermions left in band 1.
One gets 
\begin{equation}
\frac{\tilde{g}_{12}^2}{\rho_1\ln(T_c^0/T_{c1}^0)}=
\frac{\rho_2 \omega_0 (|\mu_2|-|\mu_B|)}{|\mu_2||\mu_B|}
\end{equation}
where $\mu_2=\mu_2(T_c^0)$ is the chemical potential measured 
with respect to the bottom of the band 2
and $\mu_B=\rho_2\omega_0g_{22}$ represents the bound-state energy.
In the present case
most of the fluctuation effect has been taken into account
by the formation of the bound state occurring at a very high
$T^*\sim g_{22}$. In this new physical situation
$\eta \sim \eta_1$ stays sizable and
the fluctuations will not strongly further
reduce $T_c^r$ with respect to $T_c^0$: 
$T_c\simeq T_c^r\simeq T_c^0$.
In this low doping regime
$\frac{T^*-T_c}{T^*}$ approaches its largest
values before $T_c$ vanishes.

The strong-coupling limit of our model
shares some similarities as well as some important differences
with phenomenological models of interacting 
fermions and bosons \cite{ranninger,GIL}. In particular,
we believe that
the model considered here is more suitable to describe
the crossover to the optimal and overdoped regime,
where no preformed bound states are present and the superconducting 
transition is quite similar to a standard BCS transition.


\begin{thebibliography}{9}
\bibitem{notaswave} The more
	realistic $d$-wave problem is a straightforward generalization,
	which only involves a heavier notation.
\bibitem{notaren} In the explicit calculation we assume that the 
	largest contribution to $\delta\epsilon (T)$ comes from the 
        first order selfenergy corrections to the bare Green functions 
        of the p-p bubbles. In these corrections we cutoff
        the $2D$ logarithmic divergencies by a small
	interlayer coupling.
\end{thebibliography}
\end{document}